\documentclass[useAMS,usenatbib]{mn2e}

\usepackage{times,graphicx,amssymb,natbib}

\title[Turbulent Linewidths in Self-Gravitating Discs]{Turbulent Linewidths as a Diagnostic of Self-Gravity in Protostellar Discs}
\author[D. Forgan, P. Armitage and J. Simon]{Duncan Forgan $^{1}$\thanks{E-mail:
dhf@roe.ac.uk} Philip J. Armitage$^{2,3}$ and Jacob B. Simon$^{2}$ \\
$^{1}$Scottish Universities Physics Alliance (SUPA), Institute for Astronomy, University of Edinburgh, Blackford Hill, Edinburgh, EH9 3HJ, Scotland, UK\\
$^{2}$JILA, University of Colorado and NIST, 440 UCB, Boulder, CO 80309-0440, USA \\
$^{3}$Department of Astrophysical and Planetary Sciences, University of Colorado, Boulder, USA}
\begin{document}

\date{Accepted}

\pagerange{\pageref{firstpage}--\pageref{lastpage}} \pubyear{}

\maketitle

\label{firstpage}

\begin{abstract}
We use smoothed particle hydrodynamics simulations of massive
protostellar discs to investigate the predicted broadening of
molecular lines from discs in which self-gravity is the dominant
source of angular momentum transport. The simulations include
radiative transfer, and span a range of disc-to-star mass ratios
between $M_d / M_* = 0.25$ and $M_d / M_* = 1.5$. Subtracting off the
mean azimuthal flow velocity, we compute the distribution of the
in-plane and perpendicular peculiar velocity due to large scale
structure and turbulence induced by self-gravity. For the lower mass
discs, we show that the characteristic peculiar velocities scale with
the square root of the effective turbulent viscosity parameter,
$\alpha$, as expected from local $\alpha$-disc theory.  The derived
velocities are anisotropic, with substantially larger in-plane than
perpendicular values. As the disc mass is increased, the validity of
the $\alpha$ approximation breaks down, and this is accompanied by
anomalously large in-plane broadening. There is also a high variance
due to the importance of low-$m$ spiral modes.  For low-mass discs,
the magnitude of in-plane broadening is, to leading order, equal to
the predictions from $\alpha$ disc theory and cannot constrain the
source of turbulence. However, combining our results with prior
evaluations of turbulent broadening expected in discs where the
magnetorotational instability (MRI) is active, we argue that
self-gravity may be distinguishable from the MRI in these systems if
it is possible to measure the anisotropy of the peculiar velocity
field with disc inclination. Furthermore, for large mass discs, the
dominant contribution of large-scale modes is a distinguishing
characteristic of self-gravitating turbulence versus MRI driven
turbulence.
\end{abstract}

\begin{keywords}
stars: formation, accretion, accretion discs, methods: numerical,
radiative transfer, hydrodynamics
\end{keywords}

\section{Introduction}

\noindent Protostellar discs are at the heart of both pre-main
sequence evolution and planet formation theory.  The accretion of the
disc material onto young stellar objects (YSOs) is crucial to
explaining their phenomenology.  This requires angular momentum to be
transported outwards, so that mass may be transported inwards (see
e.g. \citealt{Pringle1981a,Lodato2008,Armitage2011}). Quantitative
models of this transport process are needed to describe the evolution
of the disc's surface density profile, the protostar's accretion rate,
the initial conditions for planet formation \citep{Chiang2010}, and
the subsequent evolution of the planets in the disc. In the simplest
description, angular momentum transport is assumed to result from
turbulence within the disc, leading to an effective turbulent
viscosity \citep{Shakura_Sunyaev_73},

\begin{equation} \nu_{\rm eff} = \alpha c_s
  H. \label{eq:nu2alpha} \end{equation}

\noindent Here, $c_s$ is the sound speed, $H$ is the scale height, and 
$\alpha$ is a parameter that characterizes the efficiency of the 
transport. Constraints from observed disc lifetimes and accretion rates
\citep{Hartmann1998} are consistent with models in which $\alpha \sim
10^{-2}$.

Prostostellar disc angular momentum transport is thought to result
from two different mechanisms (depending on the location and time in
the disc's evolution). In regions of sufficiently high levels of
ionization, the magnetorotational instability (MRI) is a plausible
source of turbulence
\citep{Balbus1991,Balbus1998,Papaloizou2003}. Further out in the disc
(or early on when the disc is quite massive), the disc may be cool and
massive enough that the disc's own self-gravity is the dominant source
of turbulence \citep{Lin1987,Laughlin1994}.  This requires that the
Toomre $Q$ parameter \citep{Toomre_1964, Durisen_review},

\begin{equation} Q = \frac{c_s \kappa}{\pi G \Sigma} <
  1.5-1.7 \label{toomre_Q} \end{equation}

\noindent where $\kappa$ is the epicyclic frequency ($\kappa=\Omega$
for Keplerian discs), and $\Sigma$ is the disc surface density.  Discs
with large $Q$ will undergo radiative cooling towards lower values.
The onset of non-axisymmetric instability produces spiral structures
in the disc, heating it through shocks and viscous-scale dissipation.
Typically, self-gravitating discs will reach a self-regulated,
quasi-steady state referred to as marginal instability
\citep{Paczynski1978}, where the heating and cooling are in
approximate balance, and $Q\sim 2$.  This produces a self-sustaining
\emph{gravito-turbulence}, which can be described by an effective
$\alpha$ \citep{Gammie,Lodato_and_Rice_04}, although there are
circumstances where this approximation fails \citep{Forgan2011}. It is
also possible that both MRI and self-gravitating turbulence could be
generated at the same time, particularly in the case of layered
accretion discs \citep{Gammie1996,Armitage_et_al_01,
  Zhu2009,Martin2011}.

Determining {\em observationally} which mechanism is driving transport 
at a particular radius in a disc is difficult. Although the 
condition for the onset of self-gravity (Equation~\ref{toomre_Q}) is 
simple and well-understood, measuring $Q$ requires an absolute 
determination of the gas surface density as a function of radius. 
This is hard. Most disc mass estimates are derived from dust tracers, 
which may be systematically biased \citep{Andrews2007a}. Independent 
observational discriminants of turbulence would therefore be very valuable. 

One promising route is to detect the turbulent broadening of molecular
lines in the infrared and/or sub-mm. For local fluid turbulence,
elementary arguments \citep{Balbus1998} suggest that the
characteristic velocity perturbations will be of order $v \sim
\sqrt{\alpha} c_s \sim 0.1 c_s$. Using shearing box simulations of MHD
turbulence, \cite{Simon2011} show that this is indeed the approximate
value of velocity perturbations at the disc midplane, but that a few
scale heights above the midplane they can be as high as $0.5 c_s$, and
that a small fraction of the broadening is in fact supersonic.  While
extracting turbulent broadening for molecular species directly from
the simulations would be ideal, doing so requires an understanding of
the disc's density and temperature fields and the abundance of the
molecular lines in question.  Modelling these properties in
observations is non-trivial, as is the subsequent modelling of
radiation transport through the disc.  Projection and resolution
effects must also be considered.  Having said this, recent
observations have successfully constrained turbulent broadening.  For
example, using the CO (3-2) transition with the Sub Millimetre Array
(SMA), \citet{Hughes2011} placed an upper limit of $v < 0.1 c_s$ in
the TW Hydra system and determined a broadening value of $v \approx
0.4 c_s$ in HD 163296.  With the advent of more sensitive millimetre
instruments such as ALMA, these constraints will be significantly
improved upon.

In this paper, we present preliminary calculations of turbulent
velocities in global, three-dimensional smoothed particle
hydrodynamics simulations of self-gravitating protostellar discs with
radiative transfer \citep{intro_hybrid}. Our goal is to characterise
the turbulent velocity distribution to first-order, and identify any
qualitative features in the turbulent velocity field that might
enable the discrimination between alternate mechanisms for driving
turbulence (although we leave the extraction of synthetic line
  observations for future work). Existing studies of the MRI and of
self-gravitating turbulence suggest that significant differences may
exist. Although large-scale magnetic fields play a role in MHD
turbulence \citep{Simon2012}, the MRI can be described to a first
approximation as a local angular momentum transport mechanism.  The
same is not true of self-gravity in the limit where the disc is
massive \citep{Balbus1999,Lodato2005,Forgan2011}. With this in mind,
we seek to identify phenomenological differences which would allow us
to distinguish whether line-broadening is MHD-driven or
gravity-driven. Such a discriminant would be complementary to
estimates based upon derived disc masses, or upon direct imaging
searches for spiral structure \citep{Cossins2010a}.

\section{Method }\label{sec:Method}

\subsection{SPH and the Hybrid Radiative Transfer Approximation}

\noindent Smoothed Particle Hydrodynamics (SPH)
\citep{Lucy,Gingold_Monaghan,Monaghan_92} is a Lagrangian formalism
that represents a fluid by a distribution of particles.  Each particle
is assigned a mass, position, internal energy and velocity.  State
variables such as density and pressure are then calculated by
interpolation (see reviews by \citealt{Monaghan_92,Monaghan_05}).  In
the simulations presented here, the gas is modelled using 500,000 SPH
particles while the star is represented by a point mass particle onto
which gas particles can accrete, if they are sufficiently close and
are bound \citep{Bate_code}.

The SPH code used in this work is based on the SPH code developed by
\citet{Bate_code} which uses individual particle timesteps, and
individually variable smoothing lengths, $h_{\rm i}$, such that the
number of nearest neighbours for each particle is \(50 \pm 20\).  The
code uses a hybrid method of approximate radiative transfer
\citep{intro_hybrid}, which is built on two pre-existing radiative
algorithms: the polytropic cooling approximation devised by
\citet{Stam_2007}, and flux-limited diffusion (e.g.,
\citealt{WB_1,Mayer2007}, see \citealt{intro_hybrid} for details).
This union allows the effects of both global cooling and radiative
transport to be modelled, without imposing extra boundary conditions.

The opacity and temperature of the gas is calculated using a
non-trivial equation of state.  This accounts for the effects of
H$_{\rm 2}$ dissociation, H$^{\rm 0}$ ionisation, He$^{\rm 0}$ and
He$^{\rm +}$ ionisation, ice evaporation, dust sublimation, molecular
absorption, bound-free and free-free transitions and electron
scattering \citep{Bell_and_Lin,Boley_hydrogen,Stam_2007}.  Heating of
the disc is achieved by \(P\,dV\) work and shock heating.  Stellar
irradiation is not included as a heating source.

\subsection{Initial Disc Conditions}

\noindent The gas discs used in this work were initialised with the
$500,000$ SPH particles located between $r_{\rm in} = 10$ au and
$r_{\rm out} = 50$ au, distributed such that the initial surface
density profile is $\Sigma \propto r^{-3/2}$ and the initial
sound speed profile is $c_s \propto r^{-1/4}$.  They were first
presented in \citet{Forgan2011} to evaluate the validity of the
$\alpha$-approximation.

\begin{figure*}
\begin{center}$
\begin{array}{cc}
\includegraphics[scale = 0.45]{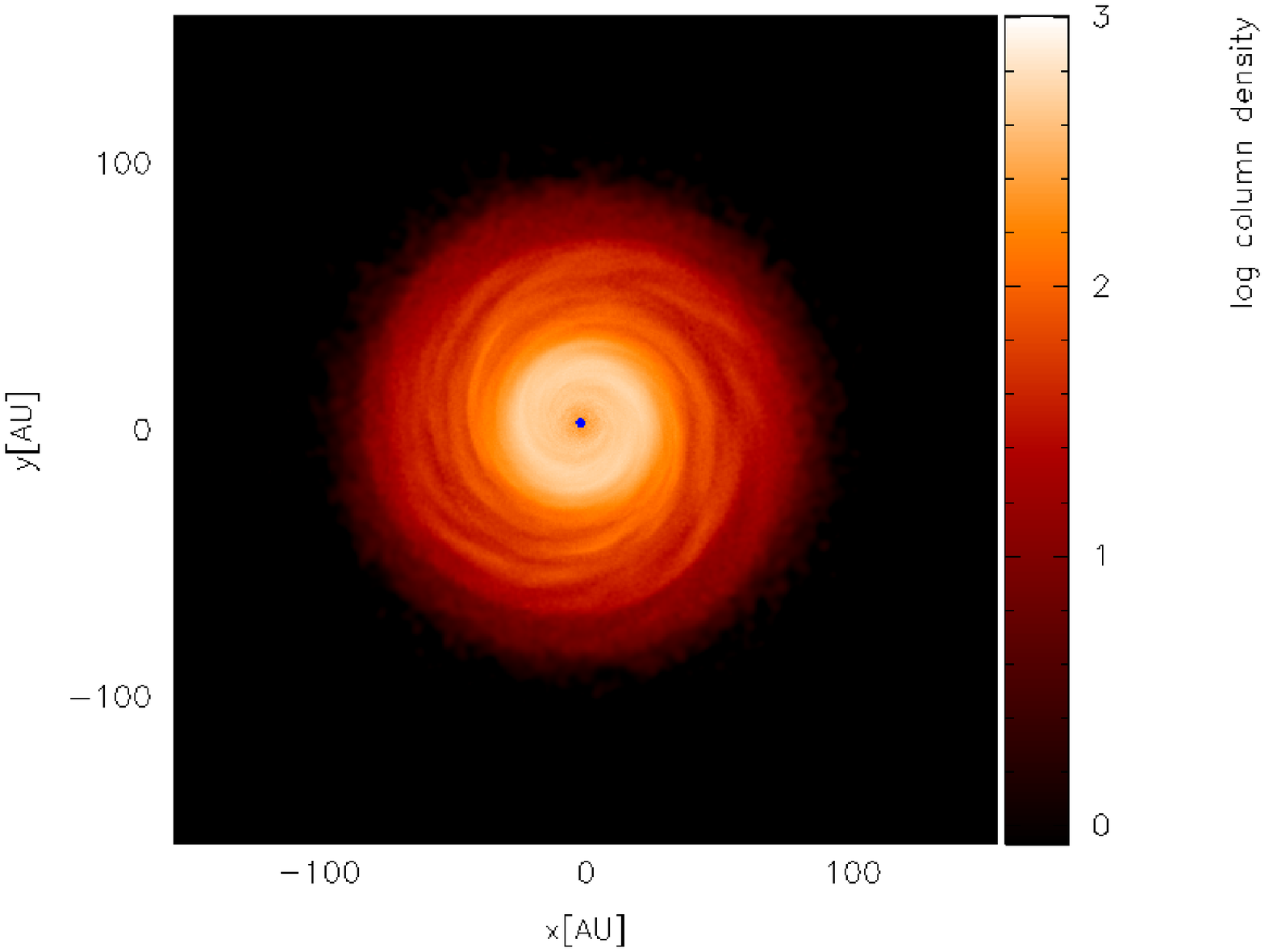} &
\includegraphics[scale = 0.45]{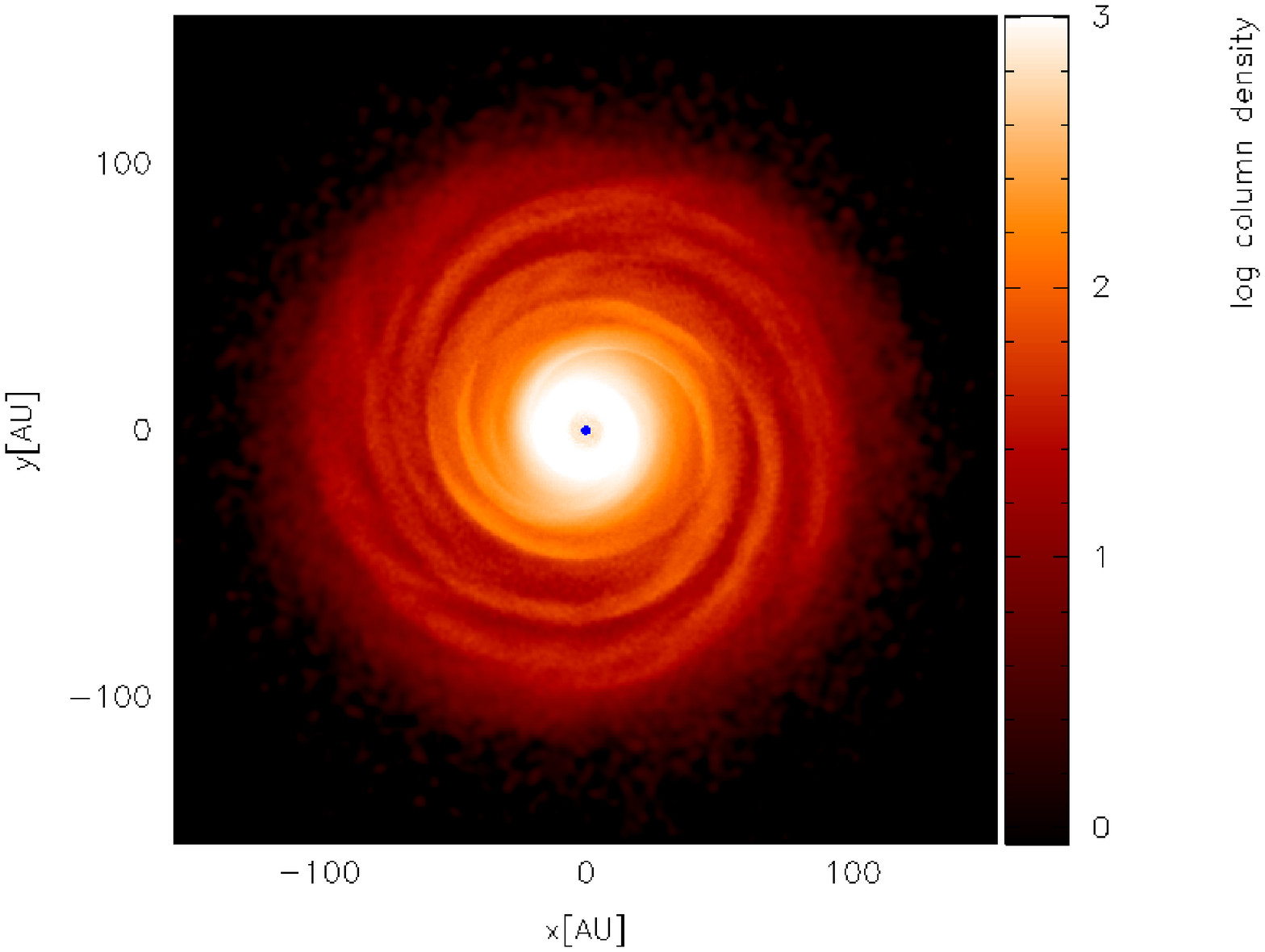} \\
\includegraphics[scale=0.45]{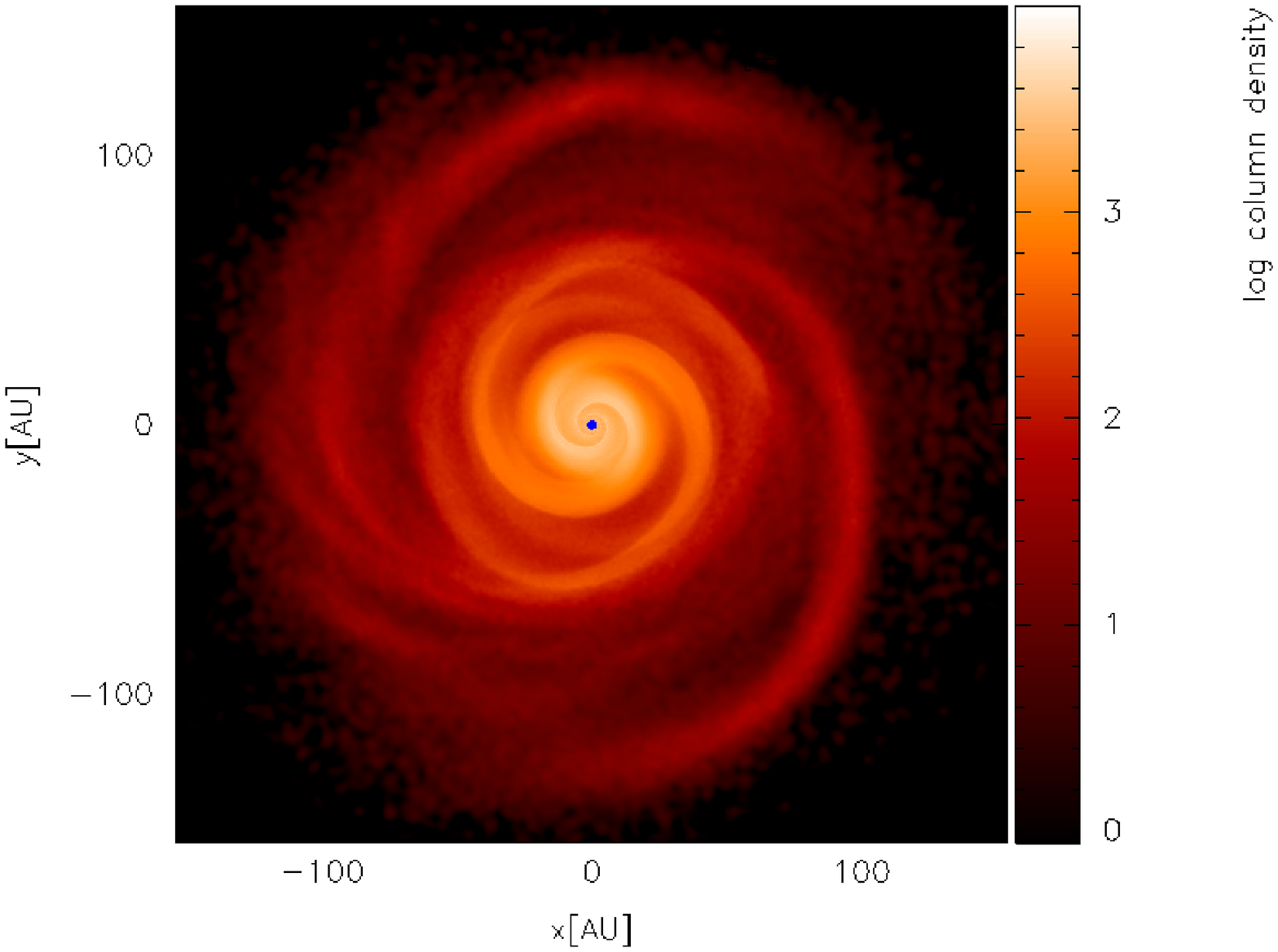} &
\includegraphics[scale=0.45]{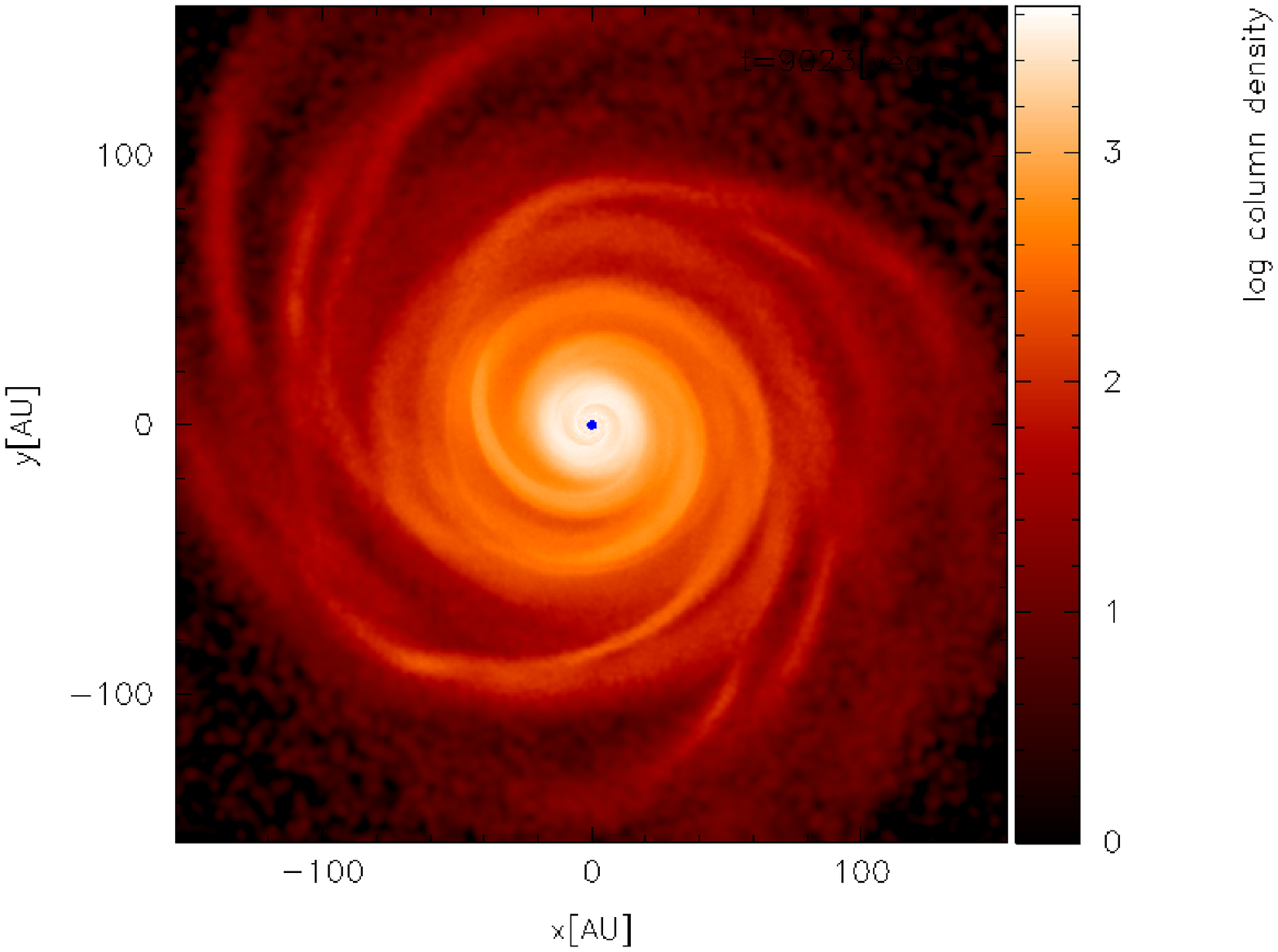}
\end{array}$
\caption{Images showing the surface density structure of Simulations 1
  (top left), 2 (top right), 3 (bottom left) \& 4 (bottom right) after
  27 outer rotation periods (ORPs).  The stellar mass in each case is
  1 $M_{\rm \odot}$, and the initial disc masses of 0.25 $M_{\rm
    \odot}$, 0.5 $M_{\rm \odot}$, 1 $M_{\rm \odot}$ and 1.5 $M_{\rm
    \odot}$ respectively. The axis ranges are shown in each figure -
  it is clear that the more massive discs exhibit higher amplitude
  spiral structures, in particular the $m=2$
  mode.  \label{fig:Ms1discs}}
\end{center}
\end{figure*}

We are primarily interested in considering quasi-steady
self-gravitating systems, rather than systems that could fragment to
form bound companions.  These initial conditions (in particular the
small disc radii) were therefore motivated by recent work suggesting
that massive discs will fragment at radii beyond $\sim 60 - 70$ au
\citep{Rafikov_05,Stam_frag,stamatellos2008a,Clarke_09,Rice_and_Armitage_09}.
This result is consistent with observations that massive discs tend to
have outer radii less than 100 au \citep{Rodriguez_et_al_05}.  A
summary of the disc parameters investigated can be found in Table
\ref{tab:params}.  Simulation 1, as the lowest mass disc in the set,
approximates local angular momentum transport; Simulation 4 exhibits
strongly non-local angular momentum transport.  The strength of
non-local effects increases as the disc mass is increased, indicated
by increasing amplitude in $m=2$ spiral modes (see Figure
\ref{fig:Ms1discs}).

\begin{table}
\centering
\begin{minipage}{140mm}
  \caption{Summary of the disc parameters investigated in this
    work.\label{tab:params}}
  \begin{tabular}{c || cc}
  \hline
  \hline
   Simulation & $M_{*}$ ($M_{\rm \odot}$) &  $M_{d}$ ($M_{\rm \odot}$)  \\  
 \hline
  1 & 1.0 & 0.25  \\
  2 & 1.0 & 0.5  \\
  3 & 1.0 & 1.0  \\
  4 & 1.0 & 1.5  \\
 \hline
  \hline
\end{tabular}
\end{minipage}
\end{table}

\subsection{Calculating line of sight velocities}

\noindent Observational constraints on turbulence (or, more generally,
on deviations from Keplerian rotation) are derived from
two-dimensional maps of the line of sight velocity field as traced by
a particular molecular line. Two distinct theoretical quantities that
can be derived from our simulations are relevant for comparing against
such measurements. One possibility is to compute the density (or
volume) weighted distribution of line of sight velocity along columns
that penetrate the disc. This metric would probe only the small-scale
turbulent velocity field. Our SPH simulations are not best-suited to
capturing this component accurately, and current observational
techniques do not resolve these scales
\citep{Hughes2011}. Alternatively, we can subtract off the
contribution to the line of sight velocity from the mean azimuthal
flow, and evaluate the distribution of the peculiar velocity field
that remains. This peculiar velocity field {\em results from} the
presence of turbulence within the disc (since by construction we have
removed the velocity of a laminar circular disc), but it need not be
the same as the local turbulent broadening.  In particular, for a
massive self-gravitating disc, significant contributions to the
peculiar velocity field arise from low-order azimuthal disc structure,
i.e. the spiral arms.  We select the second metric as it better
reflects the observables currently available.

\begin{figure}
\begin{center}
\includegraphics[scale = 0.25]{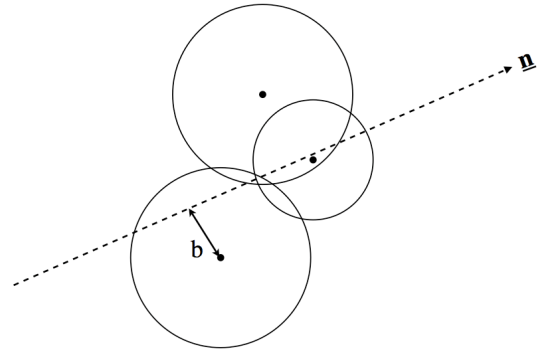}
\caption{Illustrating the raytracing method.  Particles only
  contribute to the estimated line of sight velocity if their
  smoothing volume intersects the ray.  Figure taken from
  \citet{mcsph}.  \label{fig:raytrace}}
\end{center}
\end{figure}

To compute the distribution of line of sight peculiar velocity, we
raytrace through the SPH simulation using the algorithms described in
\citet{SPHRAY} and \citet{mcsph}.  Particles only contribute to the
density and velocity fields along the ray if their smoothing
volumes\footnote{The smoothing volume is a sphere of radius $2h_i$,
  where $h_i$ is the smoothing length of particle $i$} intersect it
(see figure \ref{fig:raytrace}).  

As the smoothing volume is spherical, the task of determining
intersections is reduced to determining the closest approach of a ray
to a particle (i.e., the impact parameter $b$, as labelled in figure
\ref{fig:raytrace}), and comparing it with the particle's smoothing
length $h$.  If $b<2h$, then the particle's smoothing volume is
intersected, and its properties are used in the calculation. 

Multiple rays are drawn through a region delineated by an annulus on
the disc surface, allowing an azimuthally averaged sampling of the
velocity field while minimising the effects of particle disorder and
Poisson noise.  A total of 250,000 rays are drawn for each run,
ensuring that almost all particles contained within the annulus are
intersected by at least one ray.  We obtain the mean velocity in the
annulus - in both the vertical direction ($v_z$), and in the plane of
the disc ($v_p$) - and subtract this to obtain the peculiar velocity
field for all particles in the annulus. We then return to the list of
particles intersected by the 250,000 rays, and perform a
density-weighted average of the peculiar velocities, in the same
fashion as \citet{Simon2011}.  This is a first-order approximation to
the emissivity of the gas (i.e., we expect denser regions to radiate
more than less dense regions).

By binning the density-weighted peculiar velocity of each particle in
the list, we can construct a probability distribution of velocities
normalised by the local sound speed, $P(v/c_s)$.  These linewidth
probability distributions (LPDs) are not equivalent to any observed
line profile, but instead give the probability of observing a
particular linewidth.  Unlike \citet{Simon2011}, we do not perform
time averages of these LPDs (although how the LPD varies with time can
be seen in section \ref{sec:timedependence}).

\subsection{Limitations due to spatial resolution}
\label{limitations}

\noindent The global disc simulations that we use in this work
necessarily have lower spatial resolution than local, shearing-box
simulations. Additionally, the use of constant mass SPH resolution
elements naturally concentrates what spatial resolution we do have
toward the disc midplane, where the bulk of the mass is. These
properties of the simulations result in two limitations that we need
to be mindful of. First, we cannot study a radial extent that is too
broad, because this would result in regions where the low particle
density compromises the resolution of physical angular momentum
transport processes. Second, we cannot study the vertical dependence
of the turbulent velocity at heights where we have too few resolution
elements.

The limitation in the radial extent over which the simulation is trust
worthy derives from the use of artifical viscosity.  While required by
the SPH code used, the magnitude of artificial viscosity must be
quantified so that we know where in the disc the artificial viscosity
is likely to be lower than the effective viscosity generated by the
gravitational instabilities.  The linear term for the artificial
viscosity can be expressed as
\citep{Artymowicz1994,Murray1996,Lodato2010}:

\begin{equation} \nu_{\rm art} = \frac{1}{10} \alpha_{\rm SPH} c_{\rm
    s} h, \label{eq:nu_art}\end{equation}

\noindent where $c_{\rm s}$ is the local sound speed, $h$ is the local
SPH smoothing length, and $\alpha_{\rm SPH}$ is the linear viscosity
coefficient used by the SPH code (taken to be 0.1). We can define an
effective $\alpha$ parameter associated with the artificial viscosity
by using equation (\ref{eq:nu2alpha}) \citep{Lodato_and_Rice_04,
  Forgan2011}

\begin{equation} \nu_{\rm art} = \alpha_{\rm art} c_{\rm s}
  H, \label{eq:ss_art} \end{equation}

\noindent and hence combining equations (\ref{eq:nu_art}) and
(\ref{eq:ss_art}) gives \citep{Artymowicz1994,Murray1996,Lodato2010}

\begin{equation} \alpha_{\rm art} = \frac{1}{10} \alpha_{\rm SPH}
  \frac{h}{H}. \end{equation}

\noindent This shows that where the vertical structure is not well
resolved (i.e., $\frac{h}{H}$ is large), artificial viscosity will
dominate.  In the simulations presented here, this is typically the
case inside $\sim 10$ au (see \citealt{Forgan2011} for details), so
any data inside this region can not be used.  We therefore did not
initially populate the region inside $10$ au and although particles
will move inside $10$ au during the course of the simulations, we only
consider results outside this radius.

\begin{figure}
\begin{center}
\includegraphics[scale = 0.5]{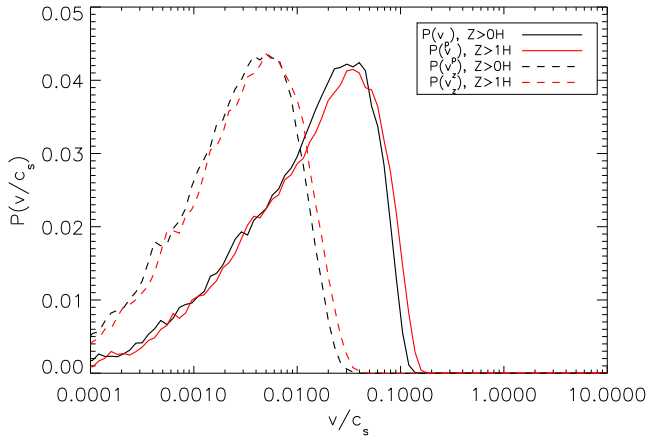}
\caption{The linewidth probability distribution in $v_p$ for
  Simulation 1, with $r=25$ au.  Planar velocities $v_p$ are plotted in
  solid lines, vertical velocities $v_z$ are plotted in dashed lines.  \label{fig:line_height}}
\end{center}
\end{figure}

There are also limitations in the vertical direction.  Unlike
\citet{Simon2011}, the simulations are not well resolved above two
scale heights, so we are unable to comment on how turbulent velocities
evolve at higher altitudes, and must be satisfied with midplane data
only.  Figure \ref{fig:line_height} shows an attempt to measure the
LPD in $v_p$ at the midplane, and at one scale height above it.  The
curves are remarkably similar, but the poor resolution of higher
altitudes forbids us from attributing this to any phenomenology of the
disc.

\section{Results}

\noindent In this section, we describe the dependence of the discs'
peculiar velocity fields on the disc mass, radius, and time from the
simulations.

\subsection{Dependence on Disc Mass}

\begin{figure}
\begin{center}
\includegraphics[scale = 0.5]{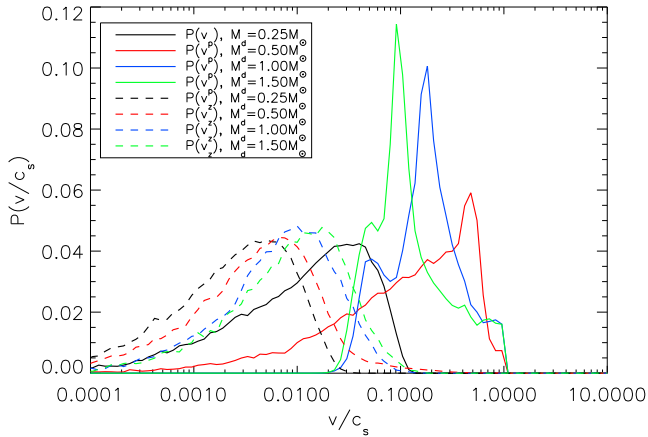}
\caption{Linewidth probability distributions for all four simulations,
  where rays are drawn through $r=25$ au.  Planar disc velocities
  $v_p$ are drawn in solid lines, and vertical velocities $v_z$ are
  drawn in dashed lines. \label{fig:mdisc}}
\end{center}
\end{figure}

\noindent As previously stated, the underlying disc mass may not be
correctly estimated by observations - can turbulent linewidths help
break this degeneracy?  Figure \ref{fig:mdisc} shows the LPDs for all
four simulations (where the annulus was placed at 25 au, with rays projected 
vertically through the disc).  

From Figure 4 in \citep{Forgan2011}, the time averaged $\alpha$ values
at 25 au
for these four discs range between $\sim 0.005- 0.01$.  Comparing the
in-plane ($v_p$) distributions to these values, we find that
Simulation 1 has a mode at $v/c_s \sim \sqrt{\alpha}$, as might be
expected from $\alpha$-disc theory.  However, increasing the disc mass
appears to break this relation, which is a symptom of the
$\alpha$-approximation itself beginning to break down in the face of
non-local angular momentum transport \citep{Balbus1999,Forgan2011}.
The LPDs change shape dramatically, developing strong peaks and
strongly asymmetric profiles.  These profiles appear to be tracing the
low-order $m=2$ spiral arms as can be seen in the lower panels of
Figure \ref{fig:Ms1discs}.  Despite this, there is a strong cutoff at
$v/c_s = 1$.  This is due to the subtraction of the average planar
motion, which for self-gravitating discs at marginal instability is
$v/c_s=1$ \citep{Cossins2008}.

The $v_z$ distributions are much more orderly, with the mode
increasing by around one order of magnitude as the disc mass is
increased by a factor of 5.  The shape of these profiles is also
reasonably constant. 

\subsection{Dependence on Disc Radius}

\noindent Having seen the striking change in LPD as the disc mass is
increased and low-$m$ spiral modes begin to significantly affect line
broadening, it is instructive to compare how LPDs change as different
disc radii are probed.

\begin{figure}
\begin{center}
\includegraphics[scale = 0.5]{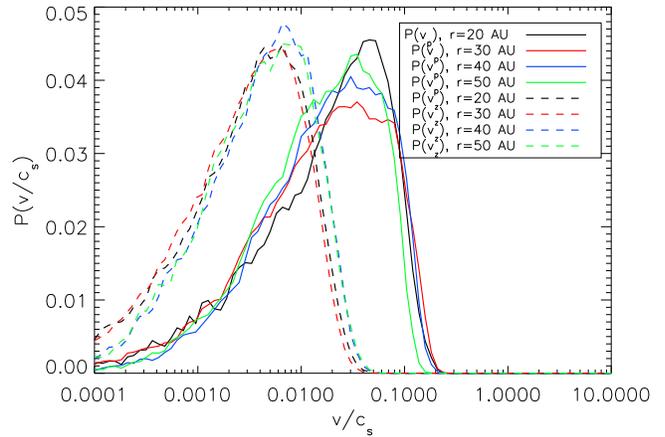}
\caption{Linewidth probability distributions for Simulation 1, where
  rays are drawn at a variety of radii.  Planar disc velocities $v_p$
  are drawn in solid lines, and vertical velocities $v_z$ are drawn in
  dashed lines.  \label{fig:rad4_Ms1Md025}}
\end{center}
\end{figure}

\noindent Figure \ref{fig:rad4_Ms1Md025} shows LPDs for Simulation 1,
drawn at a variety of disc radii.\footnote{Due to numerical
  limitations, we are restricted to a relatively small range in radii
  compared to the studies of \cite{Simon2011}; see Section
  \ref{limitations}.}  In this low mass case, the difference in peak
between planar and vertical LPDs remains constant across all radii.
Indeed, both LPDs remain very similar across all radii.  The $\alpha$
profile of Simulation 1 begins to flatten out towards a constant value
beyond $r\sim 30$ au (see \citealt{Forgan2011}), which would
presumably explain why the peaks of the LPDs tend towards similar
values (as the characteristic velocity $v/c_s \sim \sqrt{\alpha}$).

\begin{figure}
\begin{center}
\includegraphics[scale = 0.5]{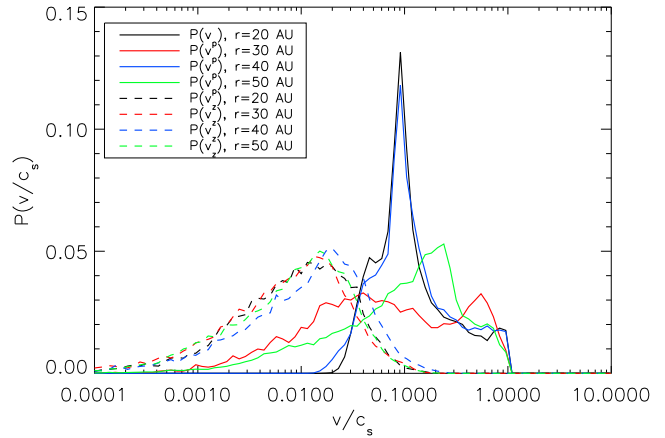}
\caption{Linewidth probability distributions for Simulation 4, where
  rays are drawn at a variety of radii at zero inclination.  Planar
  disc velocities $v_p$ are drawn in solid lines, and vertical
  velocities $v_z$ are drawn in dashed
  lines.  \label{fig:rad4_Ms1Md1_5}}
\end{center}
\end{figure}

\noindent The situation is somewhat different for Simulation 4 (Figure
\ref{fig:rad4_Ms1Md1_5}).  As the radius increases, the presence or
absence of a spiral arm inside the annulus strongly affects the
observed LPD for $v_p$.  Annuli containing an arm have a strong spike,
with a much narrower dispersion in $v/c_s$.  Annuli which do not
contain arms possess broader LPDs.  Notably, the $v_z$ distributions
remain unchanged, although this may be symptomatic of low vertical
resolution, and as such we should be wary of denoting this as a
phenomenological effect.

\subsection{Time Dependence \label{sec:timedependence}}

\noindent Finally, it is expected that linewidths should be able to
probe the inherent variability in these self-gravitating discs.  As is
shown in \citet{Forgan2011}, Simulation 1 exhibits low variability,
and Simulation 4 exhibits high variability.  These features can be
detected by calculating the LPD at a series of timesteps throughout
the disc's evolution. 

\begin{figure}
\begin{center}
\includegraphics[scale = 0.5]{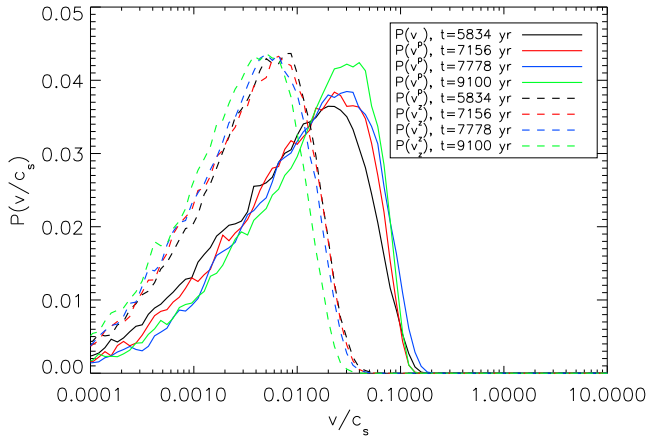}
\caption{Linewidth probability distributions as a function of time for
  Simulation 1.  Rays are drawn at $r=25$ au.  Planar disc velocities
  $v_p$ are drawn in solid lines, and vertical velocities $v_z$ are
  drawn in dashed lines.  \label{fig:time4_Ms1Md025}}
\end{center}
\end{figure}

The LPDs of Simulation 1 remain stable over several thousand years
(Figure \ref{fig:time4_Ms1Md025}), with the differences
for both $v_p$ and $v_z$ being little more than extremely small shifts
in the value of the peak.  The systematic separation between $v_p$ and
$v_z$ is maintained as a result.

\begin{figure}
\begin{center}
\includegraphics[scale = 0.5]{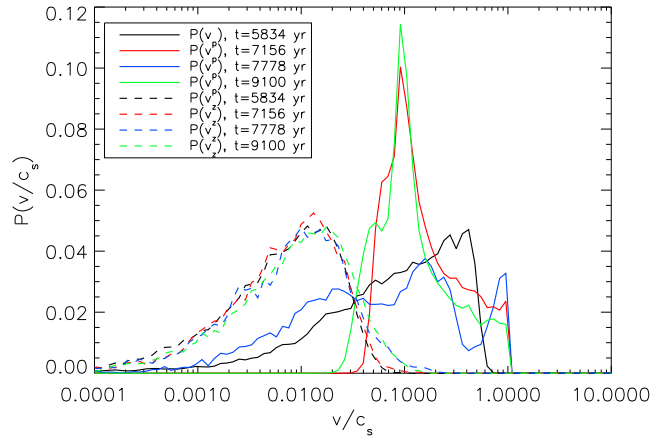}
\caption{Linewidth probability distributions as a function of time for
  Simulation 4.  Rays are drawn at $r=25$ au.  Planar disc velocities
  $v_p$ are drawn in solid lines, and vertical velocities $v_z$ are
  drawn in dashed lines.\label{fig:time4_Ms1Md1_5}}
\end{center}
\end{figure}

By contrast, the $v_p$ LPD for Simulation 4 (Figure \ref{fig:time4_Ms1Md1_5})
fluctuates much more noticeably, with both the shape of the LPD and
the amplitude of the peaks shifting over time. This is again due to
the passage of strong $m=2$ features through this particular radius.

Observers hoping to construct LPDs by observing objects should take
note.  These results would suggest that such efforts would be
successful for low mass discs, but the increased variability would
rule this method out for high mass discs.  Conversely, the simulations
show that variability can be seen in higher mass discs, even at this
comparatively low resolution.

Again, we see very little change in the $v_z$ LPD across all disc
masses, but we must be cautious as these distributions are the most
sensitive to resolution issues.

\subsection{Dependence on Numerical Resolution}

\begin{figure*}
\begin{center}$
\begin{array}{cc}
\includegraphics[scale = 0.5]{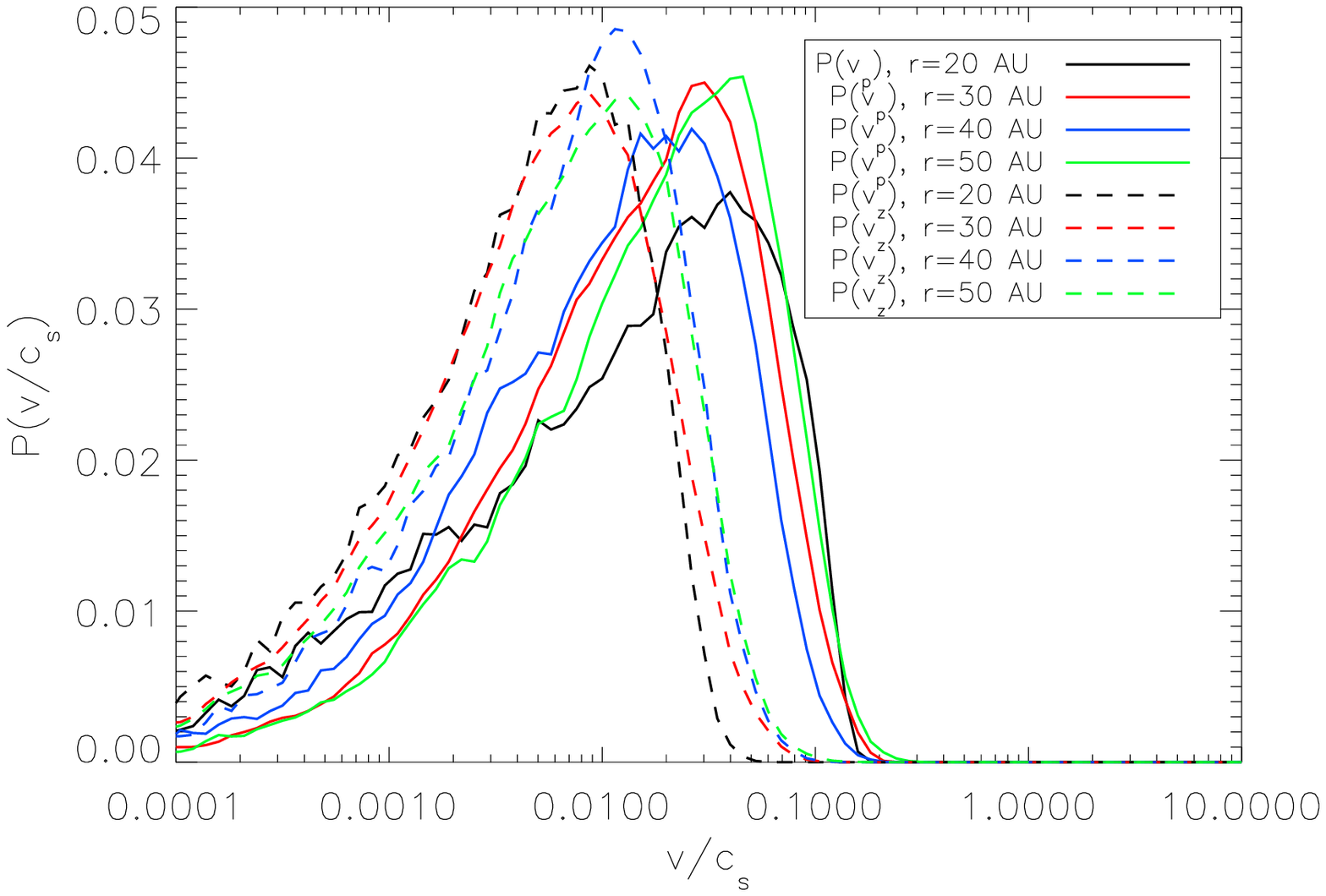} &
\includegraphics[scale= 0.5]{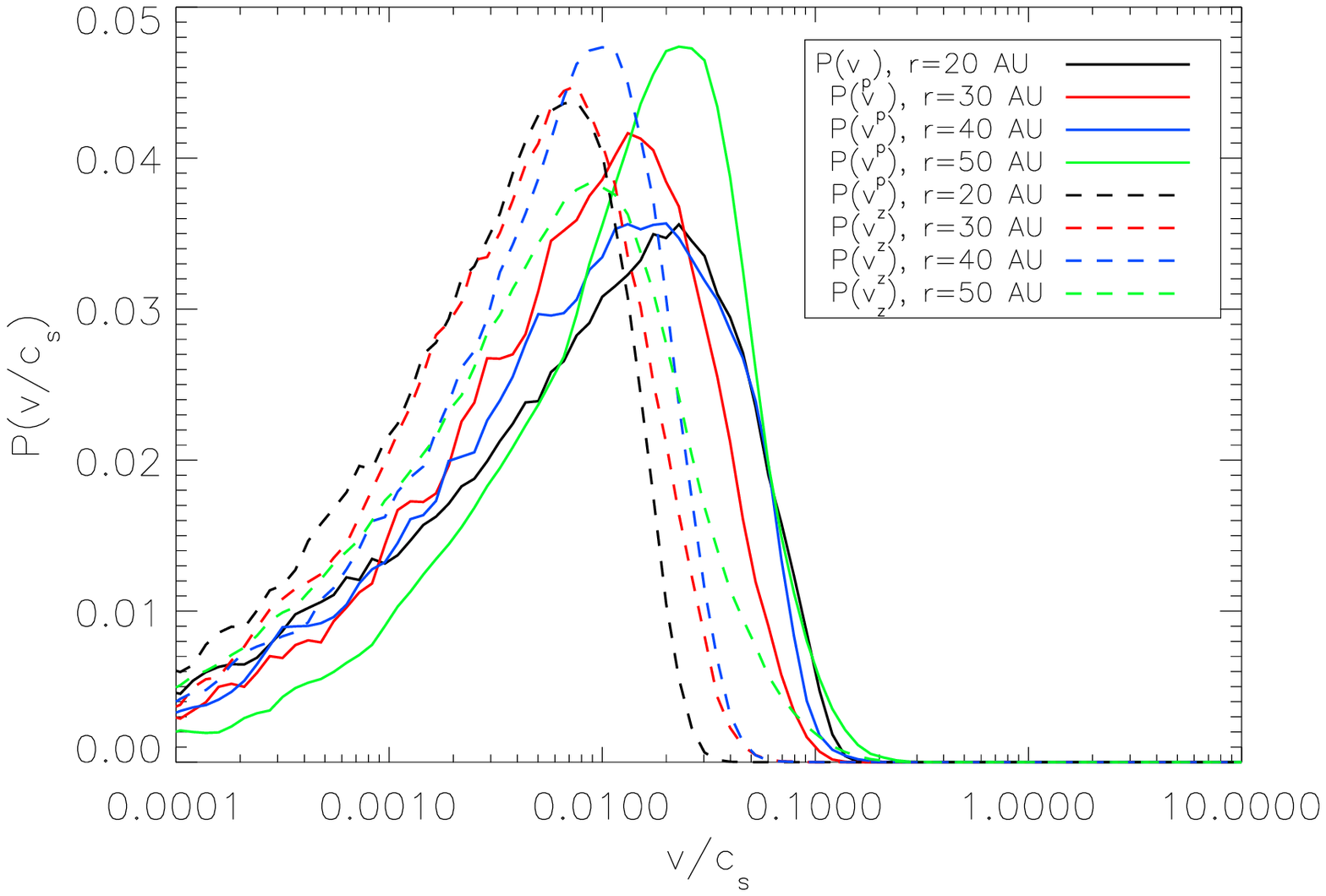} \\
\end{array}$
\caption{Linewidth probability distributions for Simulation 1 at
  different numerical resolutions, where
  rays are drawn at a variety of radii.  Planar disc velocities $v_p$
  are drawn in solid lines, and vertical velocities $v_z$ are drawn in
  dashed lines. The left panel indicates the 500,000 particle run,
  the right panel the one million particle run. \label{fig:rad4_res}}
\end{center}
\end{figure*}

\noindent Finally, we compare the effects of increasing particle
number on the resulting LPDs.  Simulation 1 was rerun with 1 million
particles (double the previous value), and LPDs were investigated at
the same instant as a function of radius.  Due to computing
constraints, the high resolution simulation was not run for the same
number of outer rotation periods as the low resolution simulation.  In Figure
\ref{fig:rad4_res} we compare the two resolutions at the same
simulation time of approximately 3500 years.

The general features of the LPDs are robust - the position of
  the peaks remains at similar values.  The $v_z$ distributions remain
  very much the same, despite increasing the vertical resolution by
  around 25\%.  Strong differences can be seen in the outer
  disc radii (blue and green curves).  This is due to extra resolution
  having a stronger effect at lower particle densities.  The inner
  disc regions still display systematic separation of $v_p$ and $v_z$,
  although the amplitude of the peaks in $v_p$ do change by around
  10\%.

\section{Discussion \& Conclusions}\label{sec:Discussion}


\noindent We have presented preliminary studies of turbulent line
broadening due to self-gravity using global smoothed particle
hydrodynamics (SPH) simulations of protostellar discs.  With radiative
transfer included in these calculations, we carried out raytracing
techniques to measure the local velocity components in the disc plane
and transverse to the disc plane (relative to the local sound speed).
From these raytracing measurements, we construct a linewidth
probability distribution (LPD) which gives the probability of
detecting lines broadened to a given width.  These simulations cannot
resolve turbulence at scales below the smoothing length, and
artificial viscosity effects will also smooth out turbulence to a
lesser extent.  That being said, we have detected some promising
diagnostics that can be compared to current MRI-driven turbulence
simulations but that will require follow-up with high resolution
studies.

First, we have found that the LPD is noticeably mass dependent, with
the maximum value of $v/c_s$ increasing towards $v/c_s = 1$, where
there is a strong cutoff.  Non-local angular momentum transport is
evident from strong peaks in the distribution in $v_p$ being caused by
low-$m$ spiral structure impinging on the disc radii being observed.
This behaviour becomes apparent when the same disc is compared at
various radii of observation.

Low mass discs which possess weaker, high $m$ spiral
structure retain similar LPDs at all radii, with a systematic
separation between the peak of the $v_z$ distribution and the peak of
the $v_p$ distribution; $v_p$ peaks near $\sim 0.04 c_s$, whereas
$v_z$ peaks near $\sim 0.005-0.01 c_s$.  

The results from this work can be compared with the work of
\cite{Simon2011} as a way to extract potential differences between the
LPDs of MRI-driven turbulence and motions due to self-gravity.  First,
comparing the actual mid-plane values of $v/c_s$, we find roughly the
same (though somewhat lower) peak $v_p/c_s$ in our low mass discs
compared to MRI-driven discs in the ideal MHD limit \citep{Simon2011}.
However, the peak values of $v_p/c_s$ in this work are slightly larger
than the dead-zone values of the 4 au simulation of \cite{Simon2011}.

The largest difference between the LPDs of low mass self-gravitating
discs and MRI-driven turbulent discs is the significant separation of
$v_z$ and $v_p$ in the self-gravitating case; this robust result is
not present in the MRI-driven simulations as $v_z$ and $v_p$ have
roughly the same LPD \citep{Simon2011}.  As such, it would appear to be
a good diagnostic for distinguishing between these two angular
momentum transport mechanisms in the limit of low disc mass.

In the high disc mass limit, the LPD fluctuates strongly in $v_p$ as
the observation radius is changed because spiral arms move in and out
of the annulus.  No such fluctuation is detected in $v_z$, but this is
possibly due to resolution effects smoothing them out.  The high time
variability exhibited by these arms is also apparent in the LPDs as
measured across time.  This is another feature that was not observed
in the simulations of \cite{Simon2011}.  The angular momentum
transport in these simulations was highly local and there were no low
$m$ features observed in the LPDs.  However, these particular
simulations were carried out in the local limit; going to larger
scales by performing global MRI simulations may introduce the presence
of low $m$ features into the LPD \citep{Beckwith2011, Simon2012}.

In conclusion, the global simulations carried out in this work suggest
that one could potentially distinguish between MRI-driven turbulence
and self-gravitational effects by observing the variation of turbulent
line width with radius (to potentially probe low $m$ structures) and
inclination angle.  We must reiterate, however, that our simulations
are quite low in resolution, and as such, important phenomenology may
not be exhibited.  In addition to this, stellar irradiation - an
important driver of line emission from the disc surface - is not
included as a radiation source.  We recommend that local simulations
be carried out in much the same manner as \citet{Simon2011}.  These
simulations would be capable of confirming the features of
gravito-turbulence seen in the low-mass discs, whose angular momentum
transport is locally determined \citep{Forgan2011}.  This includes the
systematic separation of $v_z$ and $v_p$.  Also, the increase in
resolution afforded by local simulations would allow the study of
gravito-turbulent broadening as a function of disc altitude, giving a
further means of comparison to MRI turbulence. Stellar irradiation
must also be considered in future simulations used for linewidth
studies.Finally, it is clear that more sophisticated attempts to make
synthetic observations of the linewidth broadening are required to
determine the detectability of these features using current and future
telescopes.

It is also worth noting that while theory indicates that
self-gravitating protostellar discs should exist only in the early
embedded phase of protostar formation, this does not present an
insurmountable obstacle to observing these features.  Judicious
selection of a line with high critical density for activation should
allow the midplane to be probed without contamination from the
envelope.

\section*{Acknowledgments}

\noindent Density plots were produced using SPLASH \citep{SPLASH}.
Simulations were performed using high performance computing funded by
the Scottish Universities Physics Alliance (SUPA).  DF gratefully
acknowledges support from STFC grant ST/J001422/1. PJA and JBS
acknowledge support from NASA (NNX09AB90G, NNX11AE12G) and the NSF
(0807471).

\bibliographystyle{mn2e} 
\bibliography{sph_turbvel}

\appendix

\label{lastpage}

\end{document}